\documentclass[aps,prb,twocolumn,groupedaddress,floatfix,amsmath,amssymb,amsfonts]{revtex4}

\usepackage{graphicx}

\usepackage{color}

\begin{document}

\title{Thermal phase transitions in the
attractive extended Bose-Hubbard Model with three-body constraint}

\author{Kwai-Kong Ng}
\email{kkng@thu.edu.tw} %
\affiliation{Department of Physics, Tunghai University, Taichung
40704, Taiwan}

\author{Min-Fong Yang}
\affiliation{Department of Physics, Tunghai University, Taichung
40704, Taiwan}

\date{\today}

\begin{abstract}
By means of quantum Monte Carlo simulations implemented with a
two-loop update scheme, the finite-temperature phase diagram of a
three-body constrained attractive Bose lattice gas is
investigated. The nature of the thermal phase transitions around
the dimer superfluid and the atomic superfluid is unveiled. We
find that the $Z_2$ symmetry-breaking transitions between these
two superfluid phases are of first order even at nonzero
temperatures. More interestingly, the thermal transition from the
dimer superfluid to the normal fluid is found to be consistent
with the Kosterlitz-Thouless type but giving an anomalous
universal stiffness jump. It demonstrates that this transition is
driven by unbinding of pairs of fractional vortices.
\end{abstract}


\maketitle


Since the early proposals of possible pair superfluidity in an
attractive Bose gas,~\cite{Valatin1958} the existence of this
state of matter at low temperatures has been intensively
explored.~\cite{Evans1969_Nozieres1982,Rice1988_Kagan02} 
Recent developments on the manipulation of ultracold gases may
provide direct routes in the search for the pair superfluid phase
and in probing quantum critical behaviors around this intriguing
state. In the context of Bose gases in the continuum, the
possibility of observing such a pair superfluid phase near a
Feshbach resonance has been
proposed.~\cite{Radzihovsky04_Romans04,Lee's04,Sengupta05,Radzihovsky08}
Another realization in attractive bosonic lattice gases with
three-body on-site constraint is also suggested
recently.~\cite{Diehl10}
While the discussions of this pair condensed state was originally
focused on the system of bosonic particles, similar physics can be
applied to explain some exotic phases in other physical systems.
For example, the quantum spin nematic state in some frustrated
spin systems can be understood as the condensate state of bound
magnon pairs.~\cite{Shannon06,Zhitomirsky10}

The pair superfluid phase or the dimer superfluid (DSF) phase
consists in the formation of a macroscopic coherent state made of
boson pairs.
Here we focus on the case of single-species bosons.
Contrast to the conventional atomic superfluid (ASF) with
non-vanishing order parameters $\left\langle a \right\rangle \neq
0$ and $\left\langle a^{2} \right\rangle \neq 0$ (here $a$ denotes
the boson annihilation operator), DSF is characterized by the
vanishing of atomic order parameter ($\left\langle a \right\rangle
= 0$) but nonzero pairing correlation ($\left\langle a^{2}
\right\rangle \neq 0$).
Apart from the above local order parameters, one can use
superfluid stiffness to identify the superfluid states. It is
expected that, in the DSF phase, pairs of bosons will wind
together around the system, such that only even numbers of winding
can occur. Hence the DSF phase can also be characterized by a
nonzero pair superfluid stiffness given by even winding numbers
$\rho_{\rm even} \neq 0$ \emph{and} a zero stiffness corresponding
to odd winding numbers $\rho_{\rm odd}=0$ [see
Eq.~\eqref{SF_even_odd}].~\cite{Schmidt06} Such an even-odd effect
will not appear in the ASF phase.
From the symmetry-breaking perspective, the $U(1)$ symmetry of the
global phase transformation $a \rightarrow e^{i\varphi} a$ is
completely broken in the ASF phase, while a residual $Z_2$
discrete symmetry remains in the DSF phase due to the
$\pi$-periodicity of $\langle a^{2} \rangle$.
As a result, there should exist an Ising-like quantum phase
transition between these two phases upon tuning system parameters.
Nevertheless, as pointed out in
Refs.~\onlinecite{Lee's04,Sengupta05,Radzihovsky08,Diehl10},
quantum fluctuations can turn this transition into a first-order
one due to the Coleman-Weinberg mechanism.~\cite{Coleman-Weinberg}
Although there are many theoretical investigations on the
existence of the DSF phase at zero temperature and on the nature
of the related quantum phase
transitions,~\cite{Radzihovsky04_Romans04,Lee's04,Sengupta05,Radzihovsky08,Diehl10,Lee10,1D}
reliable quantitative predictions, especially for the DSF-ASF
transition, have not yet been provided except for the
one-dimensional case.~\cite{1D} Moreover, the physics of the
thermal transitions out of the DSF phase has neither been
addressed.

\begin{figure}[tb]
\includegraphics[clip,width=0.9\columnwidth]{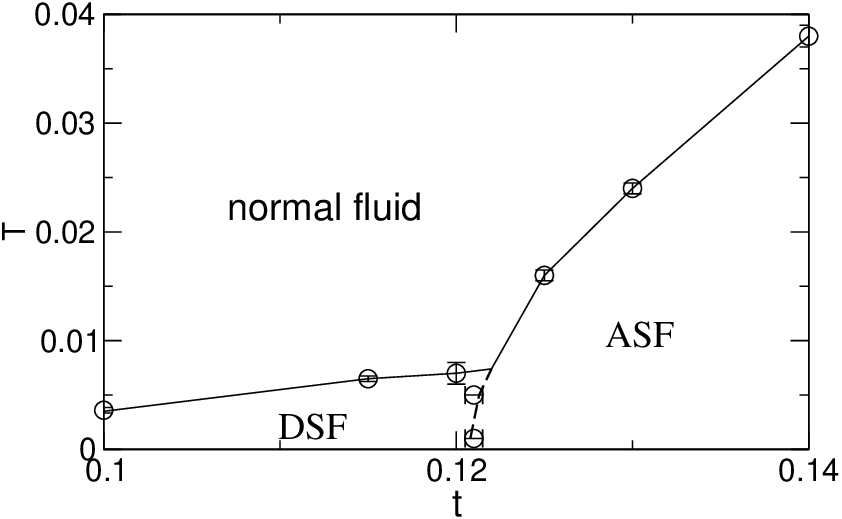}
\caption{
Finite-temperature phase diagram of a three-body constrained
attractive Bose gas on a square lattice as described in
Eq.~\eqref{eqn:H} with $|U|=1$, $V=0.25$, and $\mu=-0.55$. The
solid lines indicate the DSF-N and ASF-N transitions of continuous
KT type, and the dashed line shows the first-order DSF-ASF
transition. Lines are guide to eyes. } \label{fig:phase_diag}
\end{figure}

In the present work, the nature of the finite-temperature phase
transitions around the DSF phase in attractive bosonic lattice
gases with three-body on-site constraint~\cite{Diehl10} is
explored numerically.
We employ here the stochastic series expansion (SSE) Monte Carlo
method~\cite{SSE} generalized by allowing pair updates such that
two independent loops can merge and move together. The importance
of two-loop (or two-worm) algorithms for efficient sampling of the
DSF phase (or any paired phase) in large system sizes has been
discussed in the
literature.~\cite{Pollet06,Schmidt06,Guertler08,soyler09,Ohgoe10}
Our results for the phase diagram are summarized in
Fig.~\ref{fig:phase_diag}.
The existence of the DSF phase characterized by the even-odd
effect of the stiffness is confirmed in our simulations. Besides,
it is found that the $Z_2$ symmetry-breaking transitions between
the DSF and the ASF phases are still of first order at finite
temperatures. This indicates that the underlying Coleman-Weinberg
mechanism for the DSF-ASF transition at zero temperature is not
completely spoiled by the thermal fluctuations.
As for the transitions to normal fluids (N), both the DSF-N and
the ASF-N transitions are found to be of continuous
Kosterlitz-Thouless (KT) type.~\cite{KT} Remarkably, these two KT
transitions have distinct characters. Our data support an
anomalous value of universal stiffness jump at the KT transition
out of the DSF phase, which is \emph{four times larger} than that
out of the ASF phase. This observation clearly establishes that
the DSF-N transition is driven by the proliferation of
unconventional topological defects (half-vortices). Thus the
anomalous stiffness jump can serves as a unique signature for the
search of the DSF phase in experiments.

We consider the following extended Bose-Hubbard model with a
three-body constraint $a_i^{\dag\,3}\equiv 0$,
\begin{align}
H =& - t \sum_{\langle i,j\rangle} a_{i}^{\dagger }a_{j} %
     + \frac{U}{2} \sum_{i} n_{i}(n_{i} -1)   \nonumber \\
   & + V \sum_{\langle i,j \rangle}  n_i n_j
     - \mu \sum_{i} n_{i} \;  .
\label{eqn:H}
\end{align}
Here, $a_i (a_i^\dag)$ is the bosonic annihilation (creation)
operator at site $i$, $t$ is the hopping matrix element, $U<0$ the
on-site two-body attraction, and $\mu$ the chemical potential.
$V>0$ denotes the nearest-neighbor repulsion, 
which can come from
the dipole-dipole interactions of the dipolar bosons polarized
perpendicularly to the lattice plane by truncating it off at
the nearest-neighbor distance.
The convention $\langle i,j\rangle$ signifies a sum over
nearest-neighbor sites $i$ and $j$. The on-site constraint can
arise naturally due to large three-body loss
processes,~\cite{Daley09_Roncaglia10} and it stabilizes the
attractive bosonic system against collapse. The intriguing quantum
critical behaviors for $V=0$ at zero temperature have been
explored in Refs.~\onlinecite{Diehl10,Lee10}. Here we consider 
a finite repulsive $V$ which is found (not shown here) 
to stabilize the DSF phase and extend the region of this phase to larger $t$. 
It thus facilitates the access of the DSF phase in our simulations, 
and may also allow easy observation in the real experiments as well. 
We have also confirmed the same behaviors of the DSF for $V=0$. 
As an illustration, we choose the
parameters $V=0.25$ and $\mu=-0.55$ ($|U|\equiv 1$
 as the energy unit), so that the boson densities
are always smaller than half-filling.

To characterize different phases, several kinds of superfluid
stiffness are evaluated. In SSE, the conventional superfluid
density $\rho_{\rm s}$ at temperature $T$ is computed by measuring
the fluctuation of the winding number $W$ within the
simulations,~\cite{SSE,Pollock87}
$\rho_{\rm s} = mT\langle W^2\rangle$.
Here $m\equiv 1/2t$ is the effective mass of the bosons in a
square lattice. If a macroscopic fraction of the bosons winds
around the system, the system will give a finite $\rho_{\rm s}$.
In the ASF phase, the usual algorithm is able to let a large
number of bosons wind around the system and is known to be
efficient. However, in the DSF phase, bosons are paired into
dimers and hop together as pairs. Therefore only even winding
numbers can occur. If we define the superfluid stiffness
$\rho_{\rm even\;(odd)}$ with respect to the even (odd) winding
number $W_{\rm even\;(odd)}$ in the following way:
\begin{equation}\label{SF_even_odd}
\rho_{\rm even\;(odd)} = mT\langle (W_{\rm even\;
(odd)})^2\rangle \; ,
\end{equation}
then a clear even-odd effect, where $\rho_{\rm s}=\rho_{\rm even}
\neq 0$ and $\rho_{\rm odd}=0$, should be observed in the DSF
phase.~\cite{Schmidt06}
%
It is worth to note that there is no pair hopping term in the
Hamiltonian $H$, so that the motion of boson pairs is manifested
as a second-order effect in the single-particle hopping parameter
$t$. Within the conventional one-loop updates of the SSE
simulations, the boson numbers can only be varied by one during
each vertex update. Here in our two-loop algorithm, the pair
updates that change the boson number by two are included. As a
consequence, two independent loops can merge and move together in
such a way that effectively simulates the pair hopping in the DSF
phase. Because the accumulation of a large number of winding boson
pairs is exponentially suppressed by the short-range nature of the
usual one-loop updates, the two-loop update scheme is necessary to
improve the efficiency of the algorithm in finding the DSF phase.

\begin{figure}[tb]
\includegraphics[clip,width=0.9\columnwidth]{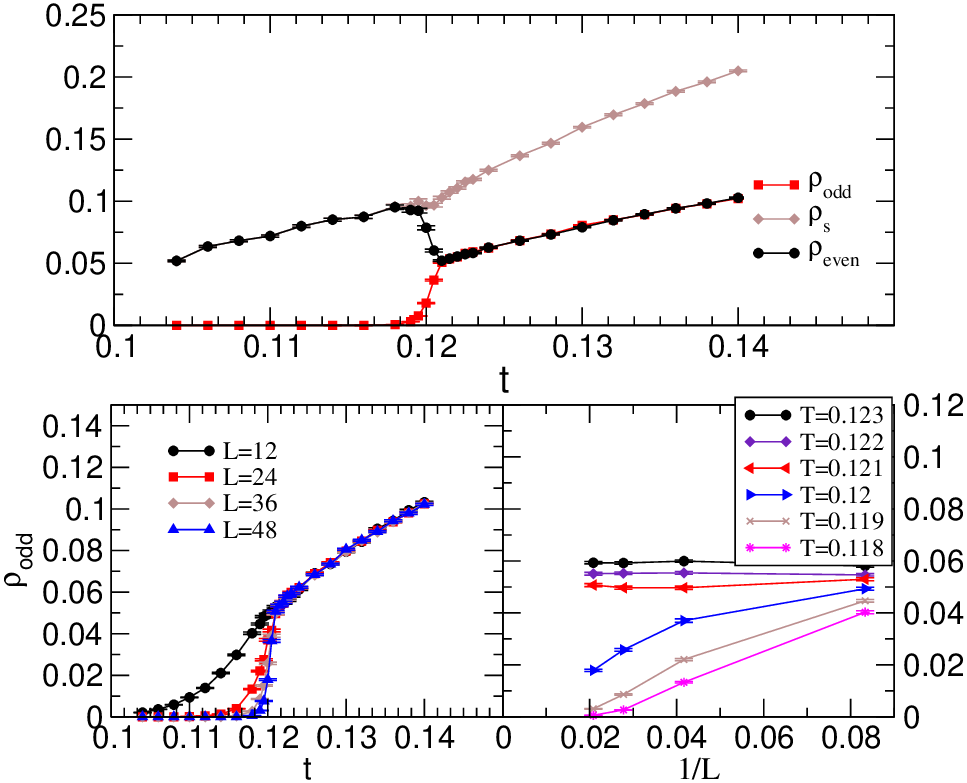}
\caption{(Color online) Upper panel: several kinds of superfluid
stiffness as functions of $t$ at $T=0.005$ with system size
$L=48$. Lower panels: data of $\rho_{\rm odd}$ for different
system sizes are shown to demonstrate the finite size effects. All
error bars are smaller than the symbol size if not shown.
 } \label{fig:DSF-ASF}
\end{figure}

As pointed out in Ref.~\onlinecite{Diehl10}, when the attraction
$U$ is strong enough, pairing correlation among bosons can be
nonzero such that the system locates in a DSF phase. Conversely,
by lowering the ratio of $|U|/t$, the ASF state can be stabilized.
Thus there is a transition separating the DSF and the ASF phases.
At zero temperature, this $Z_2$ symmetry-breaking transition is
shown to be of first order in most conditions.~\cite{Diehl10} By
performing quantum Monte Carlo simulations to study the model in
Eq.~\eqref{eqn:H} on square lattices of size $N_s=L\times L$ under
periodic boundary conditions, we find that this conclusion is
still true at low temperatures.
The results of various kinds of superfluid stiffness defined above
as functions of the hopping parameter $t$ at temperature $T=0.005$
with system size $L=48$ are shown in Fig.~\ref{fig:DSF-ASF}. It is
found that, $\rho_{\rm s}\simeq\rho_{\rm even} \neq 0$ and
$\rho_{\rm odd}=0$ for smaller hopping parameters, whereas for
larger hopping parameters $\rho_{\rm s}\simeq 2\rho_{\rm odd} \neq
0$ and there is no even-odd effect. This indicates a transition
from the DSF with $\rho_{\rm odd}=0$ to the ASF with nonzero
$\rho_{\rm odd}$. As seen from the lower panels of
Fig.~\ref{fig:DSF-ASF}, upon increasing system sizes, $\rho_{\rm
odd}$ shows an abrupt jump at $t=t_c\simeq 0.121$. It implies that
the DSF-ASF transition remains being of first order at finite
temperatures.
To further exclude the possibility of a second-order transition,
we have performed a finite-size scaling by assuming the critical
exponents of the Ising universality class. The data of different
system sizes do fail to collapse into a universal curve. This
again advocates that the DSF-ASF transition should be of first
order. Same conclusion about the nature of the DSF-ASF transition
is reached for lower temperature $T=0.001$.

As seen in Fig.~\ref{fig:phase_diag}, increasing temperature
further, either the DSF or the ASF order will be eventually
destroyed by thermal fluctuations, and will undergo a
symmetry-restoration transition to the normal-fluid state.
Before presenting the detailed analysis of these transitions,
let's begin with discussions on some general aspects of their
nature.
While both the DSF-N and the ASF-N transitions are expected to be
the continuous KT transitions, there is an essential difference
between them.
Because the DSF phase preserves the $\pi$ phase-rotation symmetry,
it can be characterized by an algebraic order in $\exp(2i\theta)$
rather than in $\exp(i\theta)$, where $\theta$ is the superfluid
phase. Therefore, the KT transition out of the DSF phase should
consist in proliferating of pairs of fractional vortices with
vorticity $\nu = \pm 1/2$, instead of the ordinary (integer)
vortices. As a result, the strength of the logarithmic interaction
of these fractional vortices will be reduced by a factor of
$\nu^2=1/4$ in comparison with that of integer vortices, and the
KT transition temperature $T_{\rm KT}$ will be decreased
substantially. The universal jump of the superfluid stiffness at
such KT transitions driven by unbinding of fractional vortices has
an anomalous value,~\cite{Korshunov02}
$\rho_s = (2/\pi\nu^2) m T_{\rm KT}$,
which amounts to
$\rho_s = 8mT_{\rm KT}/\pi$
for vorticity $\nu = \pm 1/2$. The existence of an anomalous KT
transition has been proposed in other physical
systems.~\cite{Korshunov02,Mukerjee06}

It is known that the KT transition temperatures can be determined
with good accuracy by utilizing the renormalization flow and the
universal jump of the superfluid density at the transition
point.~\cite{Weber1988,Boninsegni05} Here we generalize the data
analysis suggested in Ref.~\onlinecite{Boninsegni05} to take into
account the anomalous superfluid stiffness jump.
Consistent results can be reached by other kinds of analysis.
Define $R\equiv\pi\nu^2\rho_s/2mT$ such that $R=1$ at the KT
transition temperature $T_{\rm KT}$. It is known that the KT
renormalization group equations can be cast into an integral form,
\begin{equation}
4\ln(L_2/L_1)=\int^{R_1}_{R_2}\frac{dt}{t^2(\ln(t)-\kappa)+t}  \;,
\label{RG}
\end{equation}
where the parameter $\kappa$ is an analytic function of
temperature, and the temperature giving $\kappa=1$ corresponds to
the KT transition point. For $T<T_{\rm KT}$, a linear function in
$(T_{\rm KT}-T)$ is expected, that is, $\kappa(T) \simeq 1 + c
(T_{\rm KT}-T)$ with a positive slope $c$.~\cite{Boninsegni05}
Thus, after taking different pairs of system sizes in
Eq.~\eqref{RG} at each temperature to determine the $\kappa(T)$
curve, the location of the KT transition temperature can be
achieved by finding $\kappa(T_{\rm KT})=1$.

\begin{figure}[tb]
\includegraphics[clip,width=0.9\columnwidth]{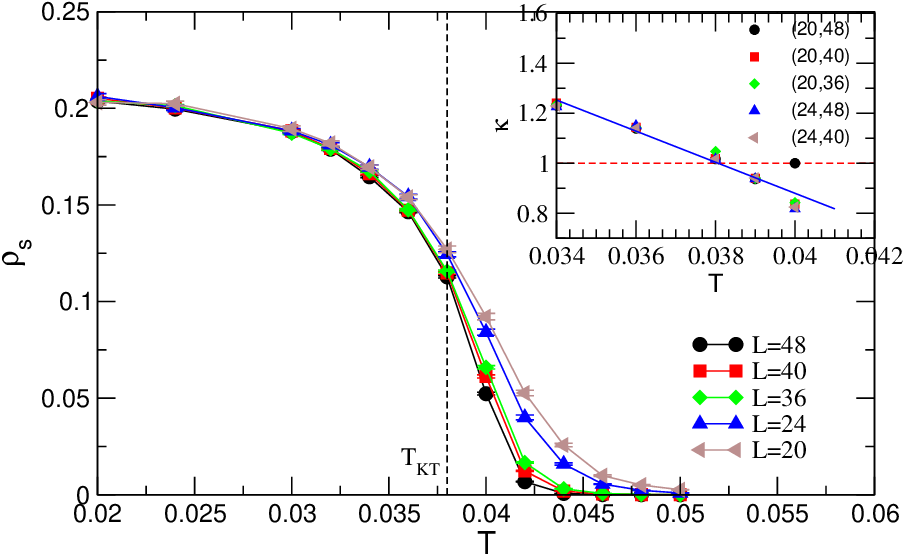}
\caption{(Color online) %
Superfluid density $\rho_s$ vs temperature $T$ with different
system sizes for $t=0.14$. The dash line denotes the location of
$T_{\rm KT}$.
Inset: solutions of the Eq.~\eqref{RG} for different pairs of
system sizes ($L_1$, $L_2$)
for
$R=\pi\nu^2\rho_s /2mT$
with $\nu=1$. The blue solid line is the linear fit
$\kappa=1+8.71(T_{\rm KT}-T)/t$
with $T_{\rm KT}\simeq 0.038$ and the red dash line is $\kappa=1$.
} \label{fig:ASF-N}
\end{figure}

Our findings of the superfluid density $\rho_s$ as a function of
temperature $T$ for $t=0.14$ are depicted in Fig.~\ref{fig:ASF-N}.
We can see the strong system size dependence characteristic to the
KT transition especially around and above the critical
temperature.
As seen from Figs.~\ref{fig:phase_diag} and \ref{fig:DSF-ASF}, the
low-temperature states at this hopping parameter belong to the ASF
phase. Thus the conventional KT transition driven by unbinding of
ordinary (integer) vortices is expected. The results of the
parameter $\kappa$ in Eq.~\eqref{RG} with $\nu=1$ for $R$ are
shown in the inset of Fig.~\ref{fig:ASF-N}. The values of
$\kappa(T)$ extracted from various pairs of system sizes clearly
collapse into a straight line around the value of $\kappa=1$. This
smooth analytic behavior of $\kappa (T)$ supports that the
transition is of the usual KT type. Besides, the KT transition
temperature is found to be $T_{\rm KT}\simeq 0.038$. Thus the
corresponding universal stiffness jump in the thermodynamic limit
is
$\rho_s = 2mT_{\rm KT}/\pi \simeq 0.086$.
%

\begin{figure}[tb]
\includegraphics[clip,width=0.9\columnwidth]{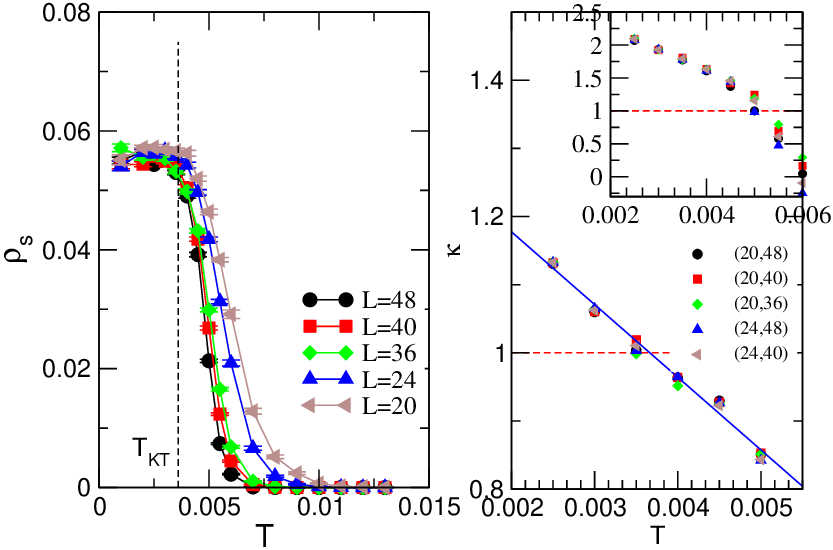}
\caption{(Color online) %
Left panel: superfluidity stiffness $\rho_s$ vs $T$ with different
system sizes for $t=0.1$. The dash line denotes the location of
$T_{\rm KT}$.
Right panel: solutions of the Eq.~\eqref{RG} for different pairs
of system sizes ($L_1$, $L_2$) for
$R=\pi\nu^2\rho_s /2mT$
with $\nu=1/2$. The blue solid line is the linear fit
$\kappa=1+10.67(T_{\rm KT}-T)/t$
with $T_{\rm KT}\simeq 0.0036$ and the red dash line is
$\kappa=1$. The inset shows the same analysis but using $\nu=1$
for $R$.  } \label{fig:DSF-N}
\end{figure}

As discussed above, the low-temperature states for small hopping
parameter $t$ can be in the DSF phase, and the anomalous KT
transition driven by unbinding of half-vortices ($\nu = \pm 1/2$)
should be observed when temperature is increased. As an
illustration, our results of $\rho_s$ vs $T$ for $t=0.1$, where
the system stays in the DSF phase at low temperatures, are shown
in the left panel of Fig.~\ref{fig:DSF-N}. Similar to the previous
data analysis but using $\nu=1/2$ for $R$, the results of
$\kappa(T)$ are presented in the right panel of
Fig.~\ref{fig:DSF-N}. Again, the data collapse is evident,
supporting that this transition is of the KT type, but originated
from the unbinding of half-vortices. For comparison, we repeat the
same analysis with $\nu=1$ for $R$, as presented in the inset of
the right panel of Fig.~\ref{fig:DSF-N}. Data collapse becomes
poor around the value of $\kappa=1$, and $\kappa(T)$ does not
behave as a linear function in $(T_{\rm KT}-T)$ for $T<T_{\rm
KT}$. These observations eliminate the possibility of the
conventional KT transition. Therefore, our findings offer strong
evidences of the existence of a half-vortex unbinding transition
out of the DSF phase and also serve as a clear signature of the
DSF phase itself. From Fig.~\ref{fig:DSF-N}, we observe that the
KT transition happens at $T_{\rm KT}\simeq 0.0036$, which is about
one order less than the value for the case of $t=0.14$. The
corresponding stiffness jump for $t=0.1$ in the thermodynamic
limit is
$\rho_s = 8mT_{\rm KT}/\pi \simeq 0.046$,
which is again less than the value for $t=0.14$.

In our simulations, the above conclusions about the nature of the
DSF-N and the ASF-N transitions apply also to the corresponding
cases with the chosen $t$'s as presented in
Fig.~\ref{fig:phase_diag}. This implies that, if it exists, the
possible region of the superfluid-to-normal-fluid transitions of
types different from the KT transition should be very narrow.


In summary, we study the DSF and the ASF phases in the extended
Bose-Hubbard model with a three-body constraint and identify the
nature of the related thermal phase transitions. The even-odd
effect in the stiffness distinguishes the DSF from the ASF phase.
The transition between these two superfluid phases is found to be
of first order. More importantly, the DSF-N transition is shown to
be driven by the topological defects of half-vortices, in great
contrast to the conventional ASF-N transition. Stimulated by the 
recent progress on detecting the KT transitions in cold atom, 
~\cite{Hung} our predictions may be verified in the near furture.

We are grateful to Y.-C. Chen for enlightening discussions and
earlier collaborations. K.-K. Ng and M.-F. Yang thank the support
from the National Science Council of Taiwan under grant NSC
97-2112-M-029-003-MY3 and NSC 99-2112-M-029-003-MY3, respectively.
After the submission of this paper we became aware of a parallel 
numerical work ~\cite{Bonnes11} for the case of $V=0$, which reaches similar 
conclusions.

\end{document}